\documentstyle{mn}

\def\mindot{{\hbox{$\buildrel'\over.$}}}

\begin{document}

\title[Intermediate-age SMC clusters]{Ages and metallicities of five intermediate-age
star clusters projected towards the Small Magellanic Cloud}

\author[Piatti et al.]{Andr\'es E. Piatti$^1$, Jo\~ao F.C. Santos, Jr.$^2$, Juan J. Clari\'a$^1$,
Eduardo Bica$^3$ \newauthor Ata Sarajedini$^4$, and Doug Geisler$^5$\\
$^1$Observatorio Astron\'omico, Laprida 854, 5000 C\'ordoba, Argentina,
andres@mail.oac.uncor.edu (AEP), claria@mail.oac.uncor.edu (JJC) \\
$^2$Departamento de F\'{\i}sica, ICEx, UFMG, CP 702, 30123-970 Belo Horizonte, MG, Brazil,
jsantos@fisica.ufmg.br \\
$^3$Universidade Federal do Rio Grande do Sul, Depto. de Astronom\'{\i}a, CP 15051, Porto Alegre,
91500-970, Brazil, bica@if.ufrgs.br \\
$^4$Astronomy Department, Van Vleck Observatory, Wesleyan University, Middletown, CT 06459,
ata@urania.astro.wesleyan.edu\\
$^5$Universidad de Concepci\'on, Departamento de F\'{\i}sica, Casilla 160-C, Concepci\'on, Chile,
doug@kukita.cfm.udec.cl}

\maketitle

\begin{abstract}

Colour-magnitude diagrams are presented for the first time for L\,32, L\,38, K\,28 (L\,43), K\,44 
(L\,68) and L\,116, which are clusters projected onto the outer parts of the Small Magellanic Cloud 
(SMC). The photometry was carried out in the Washington system $C$ and $T_1$ filters allowing the 
determination of ages by means of the magnitude difference between the red giant clump and the main 
sequence turnoff, and metallicities from the red giant branch locus. The clusters have 
ages in the range 2-6 Gyr, and metallicities between $-1.65<$[Fe/H]$<-1.10$, increasing the sample of 
intermediate-age clusters in the SMC. L\,116, the outermost cluster projected onto the SMC, is a 
foreground cluster, and somewhat closer to us than the Large Magellanic Cloud. Our results, combined 
with those for other clusters in the literature, show epochs of sudden chemical enrichment in the 
age-metallicity plane, which favour a bursting star formation history as opposede to a continuous one for
the SMC.

\end{abstract}


\section{Introduction}

It has been well known for some time that the Magellanic Clouds contain rich star clusters of all ages 
(Hodge 1960, 1961). The distribution of cluster ages, however, differs strongly between the two Clouds 
(see, e.g., Feast 1995, Olszewski et al. 1996, Westerlund\,1997). The population of recognized
genuine old clusters (with ages $\sim 12$ Gyr) in the Large Magellanic Cloud (LMC) includes possibly 
fifteen objects, 7 projected on the bar: NGC\,1835, NGC\,1898, NGC\,2005, NGC\,2019, NGC\,1916, 
NGC\,1928 and NGC\,1939, and 8 outside the bar: Reticulum, NGC\,1466, NGC\,1754, NGC\,1786, NGC\,1841, 
NGC\,2210, Hodge\,11 and NGC\,2257 (Suntzeff et al. 1992, Olsen et al. 1998, Dutra et al. 1999). 
On the contrary, although some populous metal-poor star clusters with ages between $\sim$ 5 and 9 Gyr 
are known in the Small Magellanic Cloud (SMC), only one object (NGC\,121) is known in this galaxy 
with an age of $\sim$ 12 Gyr (Stryker et al. 1985), comparable to the ages of the Galactic globular 
clusters and the oldest LMC clusters.

Regarding the intermediate-age clusters (IACs), there exists a pronounced gap in the LMC between a large
number of IACs (age $\sim$ 1-3 Gyr) and the classical old globular clusters noted above 
(Jensen et al. 1988, Da Costa 1991, van den Bergh 1991). The populous star cluster ESO\,121-SC03 
with an age of $\sim$ 9 Gyr (Mateo et al. 1986) is the only IAC in the LMC within the range 3 and 12 Gyr, 
although recent work suggests that three other populous LMC clusters (NGC\,2155, SL\,663 and NGC\,2121) may 
fall within the ``age gap'' (Sarajedini 1998). As emphasized by Olszewski et al. (1996), this gap in
the LMC cluster distribution also represents an ``abundance gap'' in that the old clusters are all
metal-poor ($<$ [Fe/H] $>$ $\sim$ -2), while the IACs are all relatively metal-rich (Olszewski et al.
1991), approaching even the present-day abundance in the LMC ($<$ [Fe/H] $>$ $\sim$ -0.5). In contrast,
the SMC is known to have a different distribution of cluster ages from the LMC (e.g., Da Costa 1991),
as it has at least six populous metal-poor star clusters with ages between $\sim$ 5 and $\sim$ 9 Gyr,
namely Lindsay\,113, Kron\,3, NGC\,339, NGC\,416, NGC\,361 and Lindsay\,1 (Mould et al. 1984,
Rich et al. 1984, Olszewski et al. 1987, Mighell et al. 1998 - hereafter MSF). Therefore, the present 
observational data 
suggest that the LMC has formed clusters in at least two different bursts, whereas the SMC has formed 
clusters more uniformly over the past 12 Gyr (although see Rich et al. 2000 for evidence favouring bursts 
in SMC cluster formation as well). The relationship between age and metallicity among the star clusters 
in both galaxies provides fundamental insight into their star formation/chemical enrichment history. Recent 
summaries of the LMC and SMC age-metallicity relations may be found in Olszewski et al. (1996), Geisler et 
al. (1997), Bica et al. (1998), MSF, Da Costa \& Hatzidimitriou (1998) and Da Costa (1999). However, 
although ages and abundances for well studied clusters in the SMC are well established, a larger sample
of SMC clusters with age/metallicity data is needed to fill out the observed cluster age - metallicity
relationship. Unlike the LMC, the SMC does not have a cluster ``age gap" which would prevent one from 
using its star clusters to learn about details of the galaxy's age-metallicity relationship. Existing SMC 
cluster age-metallicity relationships vary widely: e.g., that of Da Costa \& Hatzidimitriou (1998) 
shows continuous enrichment from the oldest to the youngest clusters and suggests the data are well fit 
by a closed box chemical evolution model, with a few anomalously metal-poor clusters at intermediate ages, 
while that of Olszewski et al. (1996) shows essentially no chemical enrichment from $\sim 10 $ Gyrs ago
until only $\sim 1-2$ Gyrs ago,  when the metallicity increased very rapidly. Clearly, more clusters are
needed to define this relationship more accurately.

The goal of the present paper is twofold: (1) to derive age and metallicity for a sample of 5 
intermediate-age cluster candidates projected towards the SMC using new CCD Washington $C,T_1$ 
photometry, and (2) to compare the cluster properties with those of their surrounding fields. The 
present data are particularly useful to improve our understanding of the age and metal-abundance 
distributions and stellar content of SMC clusters.

The selected IAC candidates are: Lindsay\,32 (L\,32) or ESO\,51-SC2, Lindsay\,38 (L\,38) or ESO\,51-SC3,
Kron\,28 (K\,28) also known as Lindsay\,43 (L\,43), Kron\,44 (K\,44) also known as Lindsay\,68 (L\,68) 
and Lindsay\,116 (L\,116) or ESO\,13-SC25, where cluster designations are from Kron (1956), Lindsay 
(1958) and Lauberts (1982). All these clusters were considered IAC candidates based on their smooth 
structure and brightness distribution of the stars, as seen on ESO/SERC Schmidt plates. Fig. 1 shows 
their positions in relation to the SMC bar. K\,28 and K\,44 are  near
the edge of the SMC main body. If the position (J2000): 00$^h$ 49$^m$ 27$^s$,
 -73$\degr$ 09$\arcmin$ 30$\arcsec$ is assumed to be the centre of the SMC bar, K\,28 is located at 
$\approx$ 1.1$\degr$ to the north, and K\,44 the same amount to the southeast. L\,32 and L\,38 at 
$\approx$ 4.2$\degr$ and 3.3$\degr$, respectively north of the bar, are among the outermost SMC clusters. 
Finally, L\,116 at 6.1$\degr$ southeast of the bar centre is the outermost projected cluster, except for 
objects located in the Bridge (Lindsay 1958, Bica \& Schmitt 1995). No colour-magnitude diagram (CMD) has 
been obtained so far for any of these SMC objects.

This paper is structured as follows: Section 2 presents the observations, while Section 3
describes the cluster and field CMDs. Section 4 focuses on ages and metallicities. Section 5 discusses 
the age-metallicity relationship in the SMC and its implication for star cluster formation. 
Finally, Section 6 deals with the conclusions of this work.

\section{Observations}

The five SMC clusters and surrounding fields were observed during four photometric nights with the Cerro
Tololo Inter-American Observatory (CTIO) 0.9m telescope in 1998 November with the Tektronix 2K \#3 CCD, 
using quad-amp readout. The scale on the chip is 0.4$\arcsec$ per pixel yielding an area covered by a 
frame of 13\mindot5$\times$13\mindot5. The integrated IRAF\footnote{IRAF is distributed by the National 
Optical Astronomy Observatories, which is operated by the Association of Universities for Research in 
Astronomy, Inc., under contract with the National Science Foundation.}-Arcon 3.3 interface for direct 
imaging was employed as the data acquisition system. A mean gain of 1.5 e$^-$/ADU and a mean readout 
noise of 4.2 e$^-$ resulted for the chosen settings. We obtained data with the Washington (Canterna 1976) 
$C$ and Kron-Cousins $R$ filters. The latter has been shown to be an efficient substitute for the standard
Washington $T_1$ filter (Geisler 1996). Exposures of 40 minutes in $C$ and 15 minutes in $R_{KC}$ were 
taken for the SMC fields. Their air-masses were always less than 1.5 and the seeing was typically 
1$\arcsec$. The observations were supplemented with nightly exposures of bias, dome- and twilight sky- 
flats to calibrate the CCD instrumental signature. Several LMC fields were also observed in the same run
using the same technique and they were presented in Piatti et al. (1999), where a detailed description 
of the data collection and reduction procedures is given. In summary, the DAOPHOT\,II/ALLSTAR stand-alone
package (Stetson 1994) was used to obtain the photometry for which the typical magnitude and colour 
errors provided by DAOPHOT\,II are shown in Fig. 2. It shows a typical trend of $T_1$ and 
$(C-T_1)$ photometric errors with $T_1$, for the cluster K\,44 and for its rich associated field. For 
the 49857 stars measured in all clusters and fields, the mean magnitude and colour errors for 
stars brighter than $T_1=19$ were 
$\sigma$($T_1$)=0.016 and $\sigma$($C-T_1$)=0.029; for stars brigther than $T_1=21$, 
$\sigma$($T_1$)=0.042 and $\sigma$($C-T_1$)=0.063. Although our photometry reaches only  slightly deeper
than the turnoff magnitudes, its quality allowed us to detect and measure the turnoff for all of them, 
which was used in our age estimates. Indeed, by using the relation between the turnoff $R$
magnitude and age according to theoretical isochrones by Bertelli et al. (1994) and by comparing it to 
our data, we concluded that we are able to define turnoffs for stellar populations as old as 6.3$\pm$1.1 Gyr 
($R$ $\approx$ 22) with an error of 0.2 in $R$. Slightly fainter turnoffs can be reached at expenses of 
larger errors. On each photometric night, a large number (typically 19-32) of 
standard stars from the list of Geisler (1996) were also observed. Care was taken to cover a wide colour
and air-mass range for these standards in order to calibrate the program stars properly. Table 1 
presents the logbook of observations of the SMC cluster fields while Fig. 3 shows the CMDs for the 
entire observed field around each cluster.
The data are available from the first author upon request.

\section{Analysis of the Colour-Magnitude Diagrams}

The relatively large size of the field of view allowed us not only to properly sample the entire extent 
of each cluster but also to sample a significant area of their surrounding field. To build cluster CMDs, 
we estimated the cluster radii by eye, selecting a limiting radius within which most of the cluster's 
light seemed to fall. The estimated radii range between 35 (14$\arcsec$) and 80 (32$\arcsec$) pixels, 
with a typical radius of 60 (24$\arcsec$) pixels. Fig. 4 shows the resulting cluster CMDs using all the
observed star within the adopted radii. All the clusters exhibit clear red giant clumps (RGCs) near $T_1$ 
$\sim$ 19 and Main Sequence (MS) turnoffs which lie roughly 0.50-0.75 magnitudes above the limit of our 
photometry, except for L\,116, whose features are more difficult to identify.

Before estimating cluster ages and metallicities, we cleaned the cluster CMDs of stars which can potentially 
belong to the foreground/background fields. We used four circular extractions placed well beyond the clusters 
and distributed throughout the observed fields. The four field regions have radii that equal half of the 
radius corresponding to the cluster in that field, so that the total field comparison area is equal to that 
of the cluster area. We then built field CMDs and counted how many stars lie in different magnitude-colour 
boxes with sizes ($\Delta$$T_1$,$\Delta$($C-T_1$)) = (0.5,0.5) mag. We then subtracted from each cluster CMD
the number of stars counted in the corresponding field CMD in each ($T_1$,$C-T1$) bin, subtracting the star 
closest to that of each field star. In Fig. 4 we represent remaining cluster stars with filled circles and
subtracted stars with open circles. In the subsequent analysis we used the former as defining the fiducial 
cluster sequences. Although the cleaned cluster CMDs may still contain some field interlopers, the CMDs
of K\,28 and K\,44 now appear to be better defined.

On the other hand, more cluster stars should also be at distances larger than the adopted radii, at least, 
as far as cluster stellar density profiles extend (see discussion below). Fig. 5 shows the resulting cluster
CMDs for circular extractions (open clusters) with radii three times larger that the adopted cluster radii, 
as well as cluster stars which define fiducial sequences (filled circles) superimposed (see Fig. 4).
As can be seen, the RGC of L\,32 includes some additional stars, the red giant branch of L\,38 is much better 
defined and the CMD of L116 has more RGC stars and a more populated MS down to fainter magnitudes. The CMDs of
K\,28 and K\,44, although containing more cluster stars, also show much greater contamination from SMC field 
stars and are presented for completeness purposes only. To estimate cluster ages and metallicities
we used these larger circular extraction CMDs weighted by the fiducial cluster stars.

Surrounding cluster field CMDs also need to be cleaned from contamination by cluster and foreground/background 
stars in order to determine their fundamental parameters and to compare properties of clusters and
associated SMC fields. Cluster extents were then delimited by adopting as field stars objects beyond 3 cluster 
radii. This criterion statistically constrains cluster star contamination in the field CMDs at a confidence
level higher than 95\%. Fig. 6 shows the resulting field CMDs plotted using all the star located between 
3$\times$(cluster radius) and CCD boundaries. The CMDs of the two inner SMC clusters of the sample (K\,28 and K\,44)
clearly reveal the main SMC field features, characterized by the mixture of young and old stellar populations. 
The most obvious features are the long MS which extends approximately 7 mags in $T_1$, the populous and
broad sub-giant branch, indicator of the evolution of stars with ages (masses) within a non negligible range, 
the RGC and the red giant branch (RGB). The RGC is somewhat elongated in $T_1$ and appeards to be populated
at brighter magnitudes by the so-called ``vertical red clump'' structure (see, e.g., Zaritsky \& Lin 1997, Gallart 
1998, Ibata et al. 1998). However, no evidence for the Vertical Structure stars seen in some LMC fields
(Piatti et al. 1999) exits.

Surrounding field CMDs are more affected by the presence of stars which belong either to the SMC or to the 
foreground Galactic field than by contamination from cluster stars. Since these Galactic field stars are
distributed over the entire field of view, we applied the statistical procedure described by MSF in order to 
remove them from the surrounding field CMDs. We assume that the Galactic field is well represented by the
surrounding field CMD of L\,116, since it has no evidence of clump or horizontal branch (HB) or turnoff of 
any kind, so that no SMC field stellar population is detected in this frame. The method is suitably designed
to clean CMDs in which the intrinsic features are well defined by many stars, as is the case for K\,28 and K\,44. 
Note that the cleaning method was only applied to L\,32 and L\,38 fields for completeness purposes,
since RGCs and MS turnoffs are clearly visible in the observed CMDs. In Fig. 7 we present probable SMC stars. 
The main features of the surrounding fields CMDs of K\,28 and K\,44 are now better defined, especially
the most evolved ones, as expected.

\section{Ages and metallicities}

\subsection{\it Star clusters}

The magnitude difference between the clump/HB and the turnoff has proved to be a useful tool for
estimating ages of IACs and old clusters as well (see Phelps et al. 1994 and references therein).
Geisler et al. (1997) calibrated this difference for the $T_1$ magnitude of the Washington system and 
applied it to a sample of LMC IACs. Following the same method, we used their calibration for estimating 
ages of our cluster sample. $\delta T_1$ magnitude differences were measured on CMDs of Fig. 5, 
assigning more weight to fiducial stars (filled circles). The cluster RGCs have an average magnitude of
$T_{1_{clump}}$ $\approx$ 19.0$\pm$0.1 mag, except for L\,116 whose RGC lies at $T_{1_{clump}}$ 
= 18.2$\pm$0.1 mag. This suggests that L\,116  is located not only several degrees from the SMC bar but 
also in front of it (see Section 5). Cluster turnoffs were more difficult to determine, mainly because 
of intrinsic dispersion and photometric errors at these faint mags. 
This was especially true for L116 which is particularly sparse. Its turnoff appears to lie either
at $T_1\sim 19.5$ or 20.2. Our preferred value is the latter, leading to an age of 2.8 Gyr; the former
value yields 1.6 Gyr. Clearly, the age for this cluster is particularly uncertain.
Photometric errors at the turnoff 
level were always $(\sigma) T_1$ $\le$ 0.15-0.20 mag. The mean $\delta T_1$ values and their errors were 
estimated from independent measurements of turnoff points and RGCs by three authors using lower and 
upper limits in order to take into account the intrinsic dispersion. The difference between maximum and 
minimum $\delta T_1$ values resulted in $\Delta(\delta T_1)$ $\approx$ 0.2 - 0.4 mags. Table 2 lists the 
resulting cluster ages computed with eq. [4] of Geisler et al. (1997). We would like to ensure that our
age scale is the same as that of MSF in which L\,1 is 9$\pm$1 Gyr old. We measured $\delta V$ = 
3.0$\pm$0.1 for L\,1 which transforms into $\delta T_1$ = 3.1$\pm$0.1 using eq. [3] of Geisler et al. 
(1997), resulting in an age of 9.5$\pm$1.0 Gyr. This value is in good agreement with that derived by 
Olzsewski et al. (1996) and Rich et al. (2000) and adopted by MSF. We did not apply any offset to our 
age scale because it is within the errors and we want to maintain consistency with the previous age 
scale of Bica et al. (1998).

As noted above, no previous CMDs exist for any of these clusters. Some age information does exist for 
K\,44, however. Elson \& Fall (1985) found K\,44 to be among the oldest SMC clusters, based on their 
{\it s} value of 47 derived from the integrated $(U-B):(B-V)$ diagram. This {\it s} value is the same 
they find for NGC\,121, generally accepted to be the oldest SMC cluster, with an age of $\sim 12$ Gyr 
(e.g. Rich et al. 2000). A search for RR Lyraes in K\,44 by Walker (1998) did not turn up any 
candidates, indicating an age $< 10$ Gyr. We find this cluster to be only a few Gyrs old. Geisler et al.
(1997) discussed the problems inherent in deriving reliable ages from integrated $UBV$ photometry of 
faint clusters in crowded fields and it appears that the Elson \& Fall estimate for K\,44 may suffer 
from this same effect.

Cluster metallicities were derived by interpolating by eye in the standard giant branches of the 
Washington system recently defined by Geisler \& Sarajedini (1999). They demonstrated  that this 
technique is three times more sensitive to metallicity than the corresponding $V-I$ technique of Da 
Costa and Armandroff (1990). To trace the standard giant branches, they used the mean loci of giant and 
subgiant branches of Galactic globular and several old open clusters with known metallicities as 
fiducial clusters. We then entered in their $M_{T_1}$ vs. $(C-T_1)_o$ diagram the T$_1$ magnitudes and 
$C-T_1$ colours for our cluster stars, previously corrected for foreground reddening and distance, and 
estimated the mean cluster metallicities. Note that Geisler \& Sarajedini (1999) derived their 
metallicity calibration for three metallicity scales -- here we use the Zinn (1985) scale. Reddening and
distance corrections were performed using the expressions $E(C-T_1) = 1.97E(B-V)$ and $M_{T_1} = T_1 + 
0.58E(B-V) - (m-M)_V$ (Geisler \& Sarajedini 1999). For the SMC clusters, we assumed an apparent 
distance modulus $(m - M)_V$ = 19.0, except for L\,116, taking into account results recently obtained by 
Cioni et al. (2000) using data extracted from the DENIS catalogue towards the Magellanic Clouds. We used 
a foreground reddening E(B-V) depending on the Galactic coordinates (Table 1) and the values from the maps
by Burstein \& Heiles (1982, hereafter BH) and Schlegel et al. (1998, hereafter SFD).  SFD produced a 
full-sky map of the Galactic dust based upon its far-infrared emission (100 $\mu$m) which allowed us to 
check the BH values. SFD have not removed the SMC so that we could take into account not only 
possible Galactic dust variations but also the internal SMC reddening, especially in the innermost SMC 
fields  K\,28 and K\,44. The BH map is based on the H\,I emission of the Galaxy. Table 3 lists the 
resulting E(B-V) values. Except for K\,28, the  cluster sample shows only small differences between the 
two colour excess estimates. The average of the BH values is 0.034$\pm$0.023, while the typical 
reddening estimated by SFD for the SMC is 0.037. Given the large discrepancy for K\,28, we will derive 
metallicities based on both reddening values. For the other clusters, we use the BH values. We recall 
that an increase of the assumed reddening by E(B-V) = 0.03 decreases the derived metallicity by 0.12 dex
(Bica et al. 1998).

Fig. 8 shows an example of a cluster CMD compared with the standard giant branches, while Table 3 lists
the resulting [Fe/H] values. Note that the metallicity for L\,116 is very uncertain given the sparcity 
of giants and the uncertainty in its distance (we used a value of 18.2 based on its RGC mag.). Since for
metallicities lower than [Fe/H] $\approx -0.5$ dex, the red giant branches were derived using Galactic 
globular clusters with ages $>$ 10 Gyr, the calibration is not directly applicable to most of our SMC 
clusters because of the noticeable effect of the age differences on broadband colours. Geisler \& 
Sarajedini (1999) found that the age effect on metallicity derivation should be small or negligible for 
clusters $>\sim 5$ Gyrs old. Bica et al. (1998) investigated the effect for younger clusters and found 
a mean offset of 0.4 dex, in the sense that the metallicities derived from the standard giant branch 
technique for younger clusters were too low compared to spectroscopically derived values. However, most 
of their sample were only 1 -- 2 Gyrs old. Lacking further details, we correct our metallicities by 
$+0.2$ dex for clusters of 3 -- 5 Gyrs and $+0.4$ dex for clusters of 1 -- 3 Gyrs. It is important to 
note that the high reddening value for K\,28 takes into account the dust along the line of sight through
the entire SMC body, and it would be appropriate for dereddening the cluster if it were behind the Small
Cloud, which is probably not the case as judged from the position of its RGC. The iron-to-hydrogen ratio
corresponding to SFD's colour excess appears in parenthesis, and for further analysis we use the value 
based on the BH reddening. The metallicity uncertainties were estimated at $\sim 0.2$ dex in all cases,
including the uncertainty in deriving the original mean value, the uncertain age correction, and reddening
and calibration errors.

\subsection{\it Surrounding fields}

Ages for surrounding fields were determined employing the same method described for clusters. Since 
fields are in most cases obviously a composite of stellar populations with different ages, we 
measured the $\delta T_1$ values for the most populous turnoffs, as done for our LMC sample (Bica et al.
1998). To assess such turnoffs along MSs of the surrounding fields of K\,28 and K\,44, we applied the 
following criterion. First, in the Galactic foreground-cleaned CMD, we defined the region corresponding
to the MS. This was accomplished by tracing a lower envelope composed of two straight lines and a 
reddest envelope shifting the lower envelope by +0.4 mag. The lines defining the lower envelope are 
given by the expressions : $T_1 = 18.0\times(C-T_1 - \alpha_1) + 28.5$ and $T_1 = 4.4\times(C-T_1 - 
\alpha_2) + 21.6$, where $\alpha_1$ and $\alpha_2$ are constants equal to 0.0 and 0.1 for K\,28 and 
L\.68, respectively. We then built MS luminosity functions by counting all the stars distributed in the 
previously delimited CMD zone and within intervals of $\Delta T_1$ = 0.5 mag. Assuming that the observed
MS is the result of the superposition of different MSs, we considered the magnitude associated with each
bin as that corresponding to a MS whose turnoff lies at that $T_1$ value. Such MS is also assumed to 
have a uniform number of stars per magnitude interval. To obtain the number of stars per bin which only 
belong to the MS turnoff in that bin, we subtracted from each interval the number of stars counted in 
the following fainter bin. Negative values reflect either that the turnoff of the fainter bin is less 
populous than that of the adjacent brighter bin or incompleteness effects due to reaching the limiting 
magnitude. The $T_1$ magnitude of the interval with the highest number of stars, after subtraction of 
fainter MS stars, was adopted as the turnoff magnitude of the most numerous stellar population of the 
surrounding cluster field. For the surrounding fields of L\,32 and  L\,38, we directly measured $\delta 
T_1$ because their turnoffs are clearly visible in  CMDs. The L\,116 surrounding field does not present 
any evidence of SMC features so that no age estimate was obtained. Table 2 lists the derived field ages. 
We point out that each field likely contains stars old enough that their turnoff is fainter than
the limit of the data. The ages that we estimate for the fields correspond to the majority of
detected stars. The more populated fields of K\,28 and K\,44 will certainly deserve detailed
modeling to explore the age structure, but the basic age of the detected stars could be inferred.

Metallicities for the surrounding cluster fields were derived in the same manner as for clusters. We did
not estimate the metallicity of the L\,116 field because of the lack of any SMC feature. To transform 
the observed ($T_1, C-T_1$) diagrams into the absolute ($M_{T_1}, (C-T_1)_o$) plane, we used the colour 
excesses E(B-V)$_{\rm BH}$ listed in Table 3. The upper MSs of the clusters and their surrounding fields
show a slight difference in colour, probably due to differences in the younger stellar population 
composition of these fields. The colour difference between the RGCs of the K\,28 and K\,44 fields is 
also less than 0.03 mag, which is in very good agreement with the cluster BH reddening difference. 
Fig. 9 shows a typical IAC field. Note that the fields generally showed a significant range in 
metallicity, amounting to $\sim$ 0.4 dex (although some of this scatter can be explained by SMC 
asymptotic giant branch stars), and that the values quoted are crude means. The same metallicity 
correction required for age effects for IAC objects were applied as for the clusters. The final metal 
abundance values are listed in Table 3, where colon means uncertain value.

\section{Discussion}

The five studied SMC clusters are spatially distributed along a curve which starts at the northwest of 
the SMC and crosses its bar practically perpendicular to the southeast. The SMC bar is approximately 
oriented in the southwest -- northeast direction. L\,32, L\,38 and K\,28 are on the northwest side of 
the bar, while K\,44 and L\,116 are located on the other side (see Fig. 1). According to the derived 
ages, the cluster sample seems to be composed of objects distributed in two age-groups with ages $\sim$ 
of 2.5 and 5.5 Gyr, respectively. Clusters in these age-groups would also appear spatially 
located in different SMC regions. The oldest clusters are preferably distributed on the northwest side 
of the bar, while the youngest ones are located on the other side. We checked this spatial age 
distribution by considering the ages of L\,113, K\,3, NGC\,339, NGC\,416, NGC\,361, L\,1 and NGC\,121 
derived by MSF, since they are on the Zinn metallicity scale and used an age scale where L\,1 is 9 Gyr, 
i.e., the same age-metallicity scale adopted in the present study. Joining our 5 clusters with these 
additional 7 clusters results in a sample of 5 and 7 objects distributed on each side of the bar. Fig. 1 
presents clusters from MSF with open triangles. The mean ages turned out to be (4.9$\pm$1.7) Gyr (n=5) 
and (6.8$\pm$2.9) Gyr (n=7) for the southeast and northwest groups, respectively. The derived mean
ages are comparable within dispersions so that star formation processes and dynamical evolution have
produced a homogeneous distribution. However the sample should be increased, and the present
observations suggest that more IACs should turn up  in future studies.

Cluster metallicities appear to follow an age-metallicity relation since our most metal-rich objects are
also the youngest clusters and the most metal-poor ones are the oldest ones of the sample. The mean 
metallicity of southeast and northwest cluster groups (MSF's clusters included) resulted in [Fe/H] = 
-1.28$\pm$0.17 (n=5) and  -1.39$\pm$0.21 (n=7), respectively. If we did not include MSF's clusters, the 
mean metallicities would be [Fe/H] = -1.1$\pm$0.10 (n=2) and -1.35$\pm$0.21 (n=3) instead of -1.3 and 
-1.4, respectively. This result suggests that the oldest clusters in the southeast group are responsible
for the most metal-poor averaged [Fe/H] value. In addition, this result also shows that there is no 
evidence of any bias, in the sense that clusters older than 5.0  Gyr should not have had their forming 
regions confined to some parts of the galaxy, but throughout the whole SMC body.

Our cluster sample considerably enlarges the number of SMC IAC and old clusters with ages and 
metallicities on the same system, thus providing us with a sufficient large number of objects with which
to investigate their age distribution. Fig. 10 shows the resulting histogram for 11 SMC clusters (7 
clusters from MSF and 4 clusters from this study). As can be seen, it would appear that clusters have 
been formed during the entire SMC lifetime, with some epochs with more intense cluster formation 
activity. In particular, Fig. 10 reveals that there could be at least two important cluster formation 
epochs at $\sim$ 3 and $\sim$ 6 Gyr qualitatively in-line with the findings of Rich et al. (2000).
The resulting absolute distance modulus for L\,116 is $(m-M)_o$ = 17.8 implying a distance from the Sun 
d$_{\odot}$ = 36 kpc. The cluster appears to be in the foreground of the SMC, and possibly also slightly 
closer than the LMC, assuming that the latter is at 50 kpc (see Bica et al. 1998). The projected distance 
from the LMC bar is $\approx$ 16$\degr$, which at the LMC distance is $\approx$ 14 kpc. This value is 
smaller than the derived cluster distance to the SMC $\approx$ 20 kpc assuming the SMC distance to the Sun 
as 63 kpc. This suggests that the cluster belongs rather to the LMC, although deeper observational
data are really required to sort out the nature of this object. Two Population II globular clusters 
considered as LMC members are as distant: Reticulum at 15.7$\degr$ and NGC\,1841 at 20.3$\degr$ convert at 
the LMC distance to 14.0 and 18.5 kpc from the LMC bar respectively. The outermost LMC IAC cluster known 
is  OHSC\,37 (Bica et al. 1998) at a projected distance from the LMC bar centre of $\approx$ 13$\degr$, or 
11.5 kpc. Population II globular clusters are expected at large distance since they may be part of an 
extended spheroid, but such far away intermediate age clusters may be rather explained by (i) cluster 
scattering during LMC-SMC interactions or (ii) star cluster formation during early LMC-SMC interactions 
in features such as bridges and tidal arms.

A comparison between the derived cluster and surrounding field ages shows that clusters are projected 
towards SMC fields generally composed of a similar stellar population; the difference between them being
$\Delta t_{\rm (cluster - field)}$ = -0.5$\pm$1.0 Gyrs. The fields of K\,28 and K\,44, besides the 
intermediate age component denoted by the clump and RGB, present a young component as revealed by the 
blue MS extending well above the clump level. This shows that the edge of the SMC main body has been active 
in star formation until quite recently. The projected linear distances from the bar centre for L\,32 and
L\,38 are 4.6 kpc and 3.6 kpc, respectively, and at such distances the SMC field population is clearly 
present (Fig. 6). However, the CMDs of these more distant fields do not show young components; at such 
distances the intermediate ages prevail. The field around L\,116 does not show evidence of an SMC 
population, the field appearing as foreground Galactic stars. We recall that at the SMC distance the 
linear separation would be 6.7 kpc. Similarly, metallicities of both clusters and their surrounding 
fields seem to be indistinguishable within the errors, with a difference of  $\Delta [Fe/H]_{\rm 
(cluster - field)}$ = 0.04$\pm$0.17 dex.

Finally, we studied the chemical enrichment of the SMC using ages and metallicities of the 7 star 
clusters observed by MSF and our present IAC sample. We included 5 young SMC clusters from Da Costa \& 
Hatzidimitriou (1998), which represent the present-day properties of the SMC, because they were also 
included by MSF in their figure 13. Fig. 11 shows the resulting age-metallicity relationship, where we 
present previously studied clusters and those discussed in this paper with open and filled circles, 
respectively. The error bars are also included. Only one cluster in our sample (L\,38) is as metal-poor 
as those of MSF. Six of   eight clusters older than 5.0 Gyr have metallicities in the range -1.7 $\le$ 
[Fe/H] $\le$ -1.4, which could suggest that the chemical enrichment would not have been very efficient 
until the last 5 Gyr. After that period, the age-metallicity relation would seem to undergo a change in 
its mean metallicity, as the metal content increases in average from [Fe/H] $\sim$ -1.5 up to -1.1 dex. 
We compare our age-metallicity relation with two theoretical models of the SMC star formation history. 
The dashed line represents a simple-closed system with continuous star formation under the assumption of
chemical homogeneity (Da Costa \& Hatzidimitriou 1998), whereas the solid line depicts the bursting star 
formation history of Pagel \& Tautvai\v{s}ien\.{e} (1998). The appearance of Fig. 11 indicates that a 
closed box continuous star formation model is a poor representation of the SMC star formation history. 
Instead, the refinement of the Pagel \& Tautvai\v{s}ien\.{e} bursting model is closer to the observed 
cluster data points. In particular, MSF suggest that the bursting model would be a better fit if the 
initial star formation epoch lasted 2 Gyr instead of 2.7 Gyr as originally assumed by the Pagel \& 
Tautvai\v{s}ien\.{e} models. Our additional cluster data points corroborate this modification by MSF.
Furthermore, we note that Da Costa (1999) has emphasized one specific weakness of the Pagel \& 
Tautvaisiene (P\&T) model. He points out that between $\sim$ 4 and $\sim$ 12 Gyr, the P\&T model 
predicts a star formation rate that is likely to be too low to produce the relatively large number of 
star clusters present in this age range. However, Da Costa (1999) also notes that this apparent 
difficulty can be resolved if the star formation rate in the model is increased to a level that is 
adequate to produce the numbers of star clusters and, at the same time, the abundance of the ISM is 
diluted by the infall of primordial or low abundance gas, which would serve to keep the overall metal 
abundance nearly constant during this time interval. Given the past interactions of the SMC with the 
LMC and the Milky Way, the possibility that the SMC was not a perfect closed-box is quite plausible.

\section{Conclusions}

New  Washington photometry  was  presented for  five clusters  (L\,32, L\,38, K\,28, K\,44, and 
L\,116) and surrounding fields projected onto the SMC  body and  outskirts. On the  basis of  their 
colour-magnitude diagrams we have determined age  and metallicity for both clusters and respective  
surrounding fields.   All  clusters turned  out  to be  of intermediate age.  One of them  (L\,116) 
probably does not  belong to the SMC, as indicated by its proximity       to the LMC. Including 
clusters and fields,  the range of ages found was 2.1  to 6.7 Gyrs and that of metallicities was 
$-1.70<$[Fe/H]$<-0.90$. The whole sample  of known intermediate-age and old  SMC clusters with
ages and metallicities determined on a uniform scale has now increased to 11.

The frequency  distribution of clusters with age  suggests two cluster formation epochs: one at 3 
and another at 6 Gyr, although more cluster observations are needed for a better definition of these 
events. On the basis of the RGC magnitude,  a distance of 36 kpc to L\,116 was obtained. Assuming  
8.5 kpc for the distance  Sun-Galactic centre, the distance  of the  cluster from  the  Galactic centre  
is $\approx$  34 kpc. The derived  deprojected distance of L\,116 to the  LMC is 18 kpc and to the  
SMC is 27\,kpc. Therefore, the cluster  is in the Galactic halo and  closer to  the LMC than  to the  
SMC. There are  10 Galactic globular  clusters farther  than $\approx$  34 kpc  from  the Galactic
centre. However,  old Galactic open clusters, more  similar to L\,116, are   not  found   that  far.   
In   the  LMC,   the  farthest   known intermediate-age cluster is OHSC\,37, at more than 10 kpc from 
the LMC bar centre. This suggests  that the properties of the intermediate-age cluster L\,116, including 
its distance, are more compatible with LMC membership.

Concerning the SMC field population, it is clear that a young stellar population component is
mixed with the intermediate-age one in the inner fields  at projected  distances of 1.2  kpc from the  
SMC centre (K\,28 and K\,44 fields). In  the outer fields associated with L\,32 and L\,38  (at  5  kpc  
and  4  kpc  respectively),  the  intermediate-age component is dominant  and the young component does  
not show up. This demonstrates that recent star formation has occurred in regions closer to the SMC body.

\section*{ACKNOWLEDGEMENTS}

The authors would like to thank the CTIO staff for their kind hospitality during the observing run. We 
would also like to acknowledge Carlos Dutra who provided us the reddening values for the cluster sample 
from the SFD et al. dust map. We also thank the referee for his valuable comments and suggestions.
A.E.P. and J.J.C. acknowledge the Argentinian institutions CONICET and Agencia Nacional
de Promoci\'on Cient\'{\i}fica y Tecnol\'ogica (ANPCyT) for their partial support. E.B. also 
acknowledges the Brazilian institutions CNPq for its support. D.G. acknowledges financial support for 
this project received from CONICYT through Fondecyt grants 1000319 and 8000002,
and from the Universidad de Concepci\'on 
through research grant No. 99.011.025.

\newpage

\begin{figure}
\caption{The position of the five studied cluster fields (filled circles) with relation to the SMC bar 
(straight line) and optical centre (cross). Clusters with ages given by Mighell et al. (1998) are also 
shown as open triangles.}
\label{fig1}
\end{figure}

\begin{figure}
\caption{Magnitude and colour photometric errors provided by DAOPHOT\,II as a function of $T_1$ for a 
rich field (K\,44) and its associated cluster. They are typical in our sample.}
\label{fig2}
\end{figure}

\begin{figure}
\caption{Washington $T_1$ vs. $C-T_1$ CMDs for all the measured stars in the cluster fields.}
\label{fig3}
\end{figure}

\begin{figure}
\caption{Washington $T_1$ vs. $C-T_1$ CMDs of star clusters. Filled circles represent probable cluster 
members and open circles removed objects (see Section 3 for details). Extraction radius in pixels is 
given in each panel.}
\label{fig4}
\end{figure}

\begin{figure}
\caption{Washington $T_1$ vs. $C-T_1$ CMDs of star clusters. Filled circles are the same as in Fig. 4 
and open circles represent stars from the larger circular extration (see Section 3 for details). The 
radius in pixels of the larger circular extraction is given in each panel.}
\label{fig5}
\end{figure}

\begin{figure}
\caption{Washington $T_1$ vs. $C-T_1$ CMDs of the surrounding fields, excluding areas of radius three 
times that of the cluster.}
\label{fig6}
\end{figure}

\begin{figure}
\caption{Washington $T_1$ vs. $C-T_1$ CMDs of the surrounding fields as in Fig. 6 statistically cleaned 
from foreground stars contamination (see Section 3 for details).}
\label{fig7}
\end{figure}

\begin{figure}
\caption{Metallicity derivation for the IAC K\,44. The cluster has been placed in the absolute $T_1$ 
magnitude - dereddened ($C-T_1$) colour plane assuming an apparent distance modulus of 19.0 and a 
reddening of E(B-V)=0.03. Standard giant branches from Geisler \& Sarajedini (1999) are marked with 
their metallicity values.}
\label{fig8}
\end{figure}

\begin{figure}
\caption{Metallicity derivation for the IAC field K\,44. The cluster field has been placed in the 
absolute $T_1$ magnitude - dereddened ($C-T_1$) colour plane assuming an apparent distance modulus of 
19.0 and a reddening of E(B-V)=0.03. Standard giant branches from Geisler \& Sarajedini (1999) are 
marked with their metallicity values.}
\label{fig9}
\end{figure}

\begin{figure}
\caption{The age distribution of SMC clusters older than 1 Gyr derived from  MFS and present cluster 
sample.}
\label{fig10}
\end{figure}

\begin{figure}
\caption{Age-metallicity relationship for star clusters in the SMC. Open circles represent data 
previously published by Da Costa \& Hatzidimitriou (1998) and Mighell et al. (1998), while filled 
circles correspond to the SMC clusters studied in this paper. Error bars are also included. The data are
compared with the closed box continuous star formation model ({\it dashed line}) computed by Da Costa \&
Hatzidimitriou (1998) for an assumed present day metallicity of -0.6 for the SMC, and the bursting model
({\it solid line}) of Pagel \& Tautvai\v{s}ien\.{e} (1998).}

\label{fig11}
\end{figure}

\newpage

\begin{table}
\caption{Observations log.}
\begin{tabular}{@{}lccccccc}\hline
Cluster & $\alpha_{2000}$ & $\delta_{2000}$ & $\ell$ & b & date & airmass & seeing \\
fields  &         (h m s) & ($\degr$ $\arcmin$ $\arcsec$) & ($\degr$) & ($\degr$)  & &  & (``) \\\hline
L\,32=ESO\,51-SC2    &   00 47 24  & -68 55 10 & 303.48 & -48.20 & 20/11/1998 & 1.28    & 1.0    \\
L\,38=ESO\,13-SC3    &   00 48 50  & -69 52 11 & 303.26 & -47.26 & 22/11/1998 & 1.32    & 1.1    \\
K\,28=L\,43          &   00 51 42  & -71 59 52 & 302.90 & -45.13 & 19/11/1998 & 1.34    & 1.0    \\
K\,44=L\,68          &   01 02 04  & -73 55 31 & 301.92 & -43.18 & 18/11/1998 & 1.40    & 1.0    \\
L\,116=ESO\,13-SC25  &   01 55 33  & -77 39 16 & 298.58 & -38.93 & 18/11/1998 & 1.49    & 1.0    \\\hline
\end{tabular}

\medskip 

Cluster identifications are from Lindsay (1958, L) and Kron (1956, K). 

The exposure times were 15 minutes for $R$ and 40 minutes for $C$.

\end{table}

\begin{table}
\caption{Ages of SMC clusters and surrounding fields.}
\begin{tabular}{@{}lcccc}\hline
Name   & $\delta T_1$ & Age$_{\rm cluster}$ & $\delta T_1$ & Age$_{\rm field}$ \\
       &   (mags)     &         (Gyr)       &    (mags)    &        (Gyr)    \\\hline
L\,32  & 2.5$\pm$0.1  & 4.8$\pm$0.5 &  2.8$\pm$0.2  & 6.7$\pm$0.8 \\
L\,38  & 2.7$\pm$0.1  & 6.0$\pm$0.6 &  2.6$\pm$0.1  & 5.4$\pm$0.2 \\
K\,28  & 1.7$\pm$0.3  & 2.1$\pm$0.5 &  2.3$\pm$0.1  & 3.7$\pm$0.4 \\     
K\,44  & 2.1$\pm$0.2  & 3.1$\pm$0.8 &  1.8$\pm$0.1  & 2.2$\pm$0.2 \\
L\,116 & 2.0$\pm$0.4::  & 2.8$\pm$1.0:: &        ---    &    ---      \\\hline
\end{tabular}

\end{table}

\begin{table}
\caption{Reddenings  and metallicities of SMC clusters and surrounding fields.}
\begin{tabular}{@{}lcccc}\hline
Name   & E(B-V)$_{\rm BH}$ & E(B-V)$_{\rm SFD}$ & [Fe/H]$_{\rm cluster}^*$ & [Fe/H]$_{\rm field}^*$ \\
\hline
L\,32  &  0.00     &      0.02           &  -1.2$\pm$0.2         &   -1.5$\pm$0.2:       \\
L\,38  &  0.02     &      0.02           &  -1.65$\pm$0.2         &   -1.7$\pm$0.2         \\
K\,28  &  0.06     &      0.16           &  -1.2(-1.45)$\pm$0.2 &    -1.2(-1.45)$\pm$0.2 \\  
K\,44  &  0.03     &      0.05           &  -1.1$\pm$0.2         &   -0.9$\pm$0.2        \\
L\,116 &  0.06     &      0.05           &  -1.1$\pm$0.2::       &    ---  \\\hline
\end{tabular}

$^*$ Metallicities were corrected by $+0.4$ and $+0.2$ for ages between 1-3 and 3-5 Gyrs. (see Section 
4 for details). 
\end{table}

\end{document}